# Optimal microwave control pulse for nuclear spin polarization and readout in dense nitrogen-vacancy ensembles in diamond

V.V. Soshenko[1,2], I.S. Cojocaru[1,2,3], S.V. Bolshedvorskii[1,2], O.R. Rubinas[1,2,3], V.N. Sorokin[1,2], A.N. Smolyaninov[2] and A.V. Akimov[1,2,3,4]

[1]*P.N. Lebedev Institute RAS, Leninsky Prospekt 53, Moscow, 119991, Russia*

[2]*LLC Sensor Spin Technologies, 121205 Nobel St. 9, Moscow, Russia*

[3]*Russian Quantum Center, Business Center "Ural", 100A Novaya St., Skolkovo, Moscow, 143025, Russia*

[4]*National University of Science and Technology MISIS, Leninsky Prospekt 4, Moscow, 119049, Russia*

email: *a.akimov@rqc.ru*

## I. ABSTRACT

Nitrogen-vacancy centers possessing nuclear spins are promising candidates for a novel nuclear spin gyroscope. Preparation of a nuclear spin state is a crucial step to implement a sensor that utilizes a nuclear spin. In a low magnetic field, such a preparation utilizes population transfer, from polarized electronic spin to nuclear spin, using microwave pulses. The use of the narrowband microwave pulse proposed earlier is inefficient when magnetic transitions are not well resolved, particularly when applied to diamond with a natural abundance of carbon atoms or dense ensembles of nitrogen-vacancy centers. In this study, the authors performed optimization of the pulse shape for 3 relatively easily accessible pulse shapes. The optimization was done for a range of magnetic transition linewidths, corresponding to the practically important range of nitrogen concentrations (5–50 ppm). It was found that, while at low nitrogen concentrations, optimized pulse added very little to simple square shape pulse, and in the case of dense nitrogen-vacancy ensembles, with a rather wide magnetic transition width of 1.5 MHz optimal pulses, a factor of 15% improvement in the population of the target state was observed.

## II. INTRODUCTION

The nitrogen-vacancy color center in diamond possessing optically detected magnetic resonance (ODMR) has proved to be an interesting platform for sensing applications. In particular, magnetic field sensing, both of high sensitivity [1–4] and high spatial resolution [5–9], was conducted, measurements of electric field were performed [10], and biocompatible temperature measurements were presented [11–13]. Most recently, rotation sensing was also conducted [14–17]. In contrast to the majority of magnetometry, electrometry, and temperature measurements, rotation measurements are heavily dependent on the utilization of nuclear spin.

Preparation of high purity nuclear spin states in nitrogen-vacancy color centers is crucial for such applications, as well as for memory-assisted magnetometry [1] as it will affect the sensor signal-to-noise ratio. Well-known techniques to polarize nitrogen nuclear spin are excited state level anticrossing [18] and dynamical nuclear polarization (DNP) [19–22]. The former is realized through optical pumping but restricts usage to a range of magnetic fields around 50 mT and field alignment to several degrees. The latter works at an arbitrary magnetic field [23] but requires a more complex pulse sequence to transfer the prepared state from electronic to nuclear spin.

Population transfer at a low field for the $\text{NV}^-$ center was initially demonstrated in the study [23]. Isotopically purified diamond samples with narrow OMDR linewidth were utilized in order to easily address nuclear-spin selective transitions. The concentration of $\text{NV}^-$ centers was also relatively small (< 1 ppm corresponding to $T_2 = 300\,\mu\text{s}$ [24]) to restrict line broadening. In later studies, laser excitation was optimized to minimize the leak of nuclear populations due to hyperfine interaction [25–27].

On the other side, denser $NV^-$ center ensembles could provide a higher optical signal output in a more compact size. Line broadening due to nitrogen [24] also renders isotope purification useless. Diamonds with a natural abundance of carbon-12 (98.9%) are more affordable and accessible for mass $NV^-$-based quantum sensor production. Moreover, higher $NV^-$ concentration means higher donor nitrogen concentration and thus a higher $NV^-/NV^0$ charge state ratio leading to a further increase in the signal-to-noise ratio of the $NV^-$ quantum sensor. However, to utilize dense $NV^-$ ensembles, band selective excitation should be optimized to provide practical initial nuclear state purity.

In the case of a dense ensemble, the current study optimized the DNP protocol, estimated bounds for nuclear polarization, and provided means for nuclear spin readout.

## III. CALCULATIONS

The $NV^-$ center is formed by one substitutional nitrogen in diamond accompanied by a vacancy instead of one of the nearby carbon atoms [28, 29]. The direction from the nitrogen atom to the vacancy forms the axis of the center. The ground state-level structure of the center is presented in Figure 1a. The spin magnitude is 1 for both electron and nuclear spin of $^{14}N$, which is the most abundant isotope of nitrogen (99.6% natural abundance). Spin-spin interaction in the ground state of the $NV^-$ center leads to splitting [29] of 2.87 GHz between components having $m_S = 0$ projection of an electron spin of the axis of the $NV^-$ center, while energy levels with $m_S = \pm 1$ are degenerate in the absence of external fields and strain. In the presence of the magnetic field, nevertheless, the degeneracy is lifted and energy levels with $m_S = -1$ and $m_S = +1$ split due to the Zeeman effect. Further calculation will assume the presence of such a small magnetic field (in a range of 10–100 G), which is typically used in experiments if level anticrossing is not desired [3, 14, 30].

The presence of the nuclear spin split each of the levels with a define projection on the $NV^-$ axis into 3 components, each with a defined component of the nuclear spin onto the $NV^-$ axis $m_I$ as it is shown in

Figure 1a. In this case, when the magnetic field is aligned along the $NV^-$ axis, the allowed magnetic dipole transition requires a total change in the spin component onto the $NV^-$ axis to be $\pm 1$.

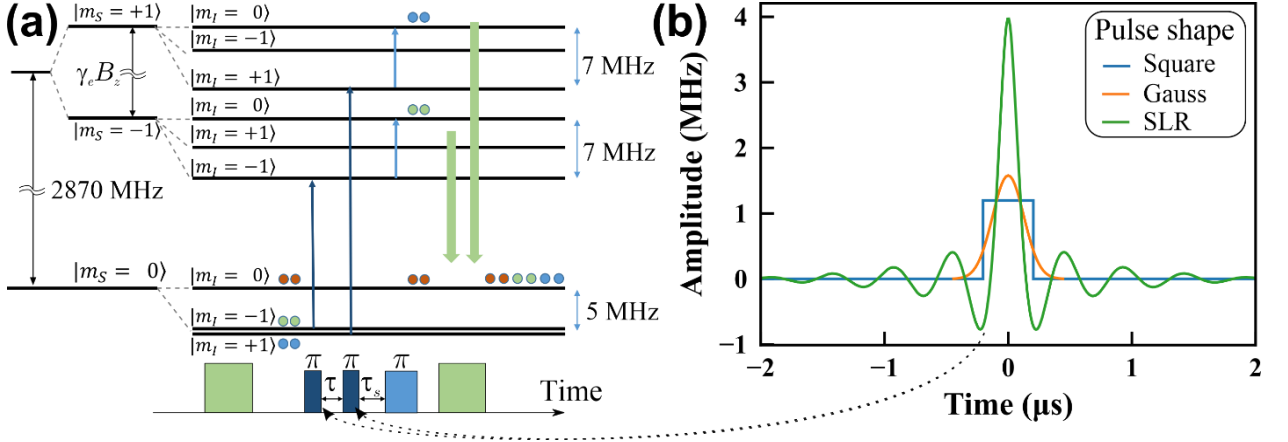

Figure 1 a) Level scheme of $NV^-$ center ground state and dynamic nuclear spin polarization protocol. b) Optimized pulse shapes for different types of pulses for the $NV^-$ center with FWHM 0.5 MHz.

To model the interaction of the $NV^-$ center with an external field, a 3*3-level model of the $NV^-$ center ground state with states defined by an electronic ($S=1$) and nuclear ($I=1$) spin projection on the $NV^-$ axis was used. Let us consider the following Hamiltonian $H_0$ for the $NV^-$ ground state:

$$\frac{H_0}{h} = D\hat{S}_z^2 + \gamma_e B_z \hat{S}_z - \gamma_n B_z \hat{I}_z + Q\hat{I}_z^2 + A_\parallel \hat{S}_z \hat{I}_z + A_\perp (\hat{S}_x \hat{I}_x + \hat{S}_y \hat{I}_y), \quad (1)$$

where $h$ is the Plank constant, $D$ is the spin-spin interaction, $\gamma_e$ is the electronic spin gyromagnetic ratio, $\gamma_n$ is the gyromagnetic ratio of the nitrogen nuclear spin, $B$ is the external magnetic field, $Q$ is quadrupole splitting, $A_\parallel$, $A_\perp$ are the longitudinal and transverse components of the hyperfine interaction tensor, and the $z$ axis is selected as an $NV^-$ axis.

To excite transitions between magnetic substates, the oscillating magnetic field $B_x(t)$ transverse to the $NV^-$ axis should be applied, leading to the time-dependent Hamiltonian:

$$\frac{H_{MW}(t)}{h} = \gamma_e B_x(t) S_x \quad (2)$$

For this study, the authors selected a specific DNP protocol of nuclear polarization, indicated in Figure 1a. For simplicity of the further simulation, the times $\tau$ and $\tau_s$ were chosen to be 0. This choice made it possible to focus on the effect of the pulse shape rather than sequence optimization. In the simulation, the authors focused their attention on the magnetically allowed transitions $|m_s=0; m_I=-1\rangle \leftrightarrow |m_s=-1; m_I=-1\rangle$ and $|m_s=0; m_I=+1\rangle \leftrightarrow |m_s=+1; m_I=+1\rangle$ which are in the two first transitions of the sequence. Two microwave pulses, shown in Figure 1a, should be tuned to the frequencies $f_{-1}=(D-\gamma_e B_z + A_\parallel)$ and $f_{+1}=(D+\gamma_e B_z + A_\parallel)$, respectively. To simplify the solution of the time-dependent master equation, which will include $H_{MW}(t)$, the authors made rotating wave approximation, connected to the oscillating field picture. Let us first assume that $B_z \gg \max_t(|B_x(t)|)$, so that the pulse applied to $|m_s=0\rangle \leftrightarrow |m_s=-1\rangle$ does not change the part of the wavefunction with $m_s=+1$ electronic spin projection and vice versa. Therefore, the rotating wave approximation is made into a rotating system, governed by the following Hamiltonian:

$$H_{RWA} = D\hat{S}_z^2 + \gamma_e B_z \hat{S}_z + Q\hat{I}_z^2 - \gamma_n B_z \hat{I}_z + (\delta_{-1} + A_\parallel)\hat{P}_{-1} + (\delta_{+1} + A_\parallel)\hat{P}_{+1},$$
$$\hat{P}_{-1} = |m_s=-1\rangle\langle m_s=-1|, \quad (3)$$
$$\hat{P}_{+1} = |m_s=+1\rangle\langle m_s=+1|,$$

where $\delta_{-1}, \delta_{+1}$ – detunings of microwave pulse frequency from the exact transition frequency for transitions $|m_s=0; m_I=-1\rangle \leftrightarrow |m_s=-1; m_I=-1\rangle$ and $|m_s=0; m_I=+1\rangle \leftrightarrow |m_s=+1; m_I=+1\rangle$, respectively.

An oscillating microwave field for the ease of solution is represented as:

$$B_x(t) = I_{-1}(t)\sin(2\pi[f_{-1}+\delta_{-1}]t) + I_{+1}(t)\sin(2\pi[f_{+1}+\delta_{+1}]t), \quad (4)$$

where $I_{-1}(t), I_{+1}(t)$ – envelope waveform of excitation pulses. Pulse shapes, used for the current study, could be implemented by amplitude modulation, so that quadrature parts in the equation are omitted. Applying rotating wave approximation for Hamiltonians $H_0, H_{MW}$ will provide:

$$\frac{H'_0}{h} = A_\parallel \hat{S}_z \hat{I}_z - \left(\delta_{-1} + A_\parallel\right)\hat{P}_{-1} - \left(\delta_{+1} + A_\parallel\right)\hat{P}_{+1}$$

$$\frac{H'_{MW}}{h} = \gamma_e \frac{1}{2\sqrt{2}}\left(-\hat{\lambda}_2 I_{-1}(t) + \hat{\lambda}_7 I_{+1}(t)\right) \otimes \hat{E} \quad ,\quad (5)$$

$$\hat{\lambda}_2 = -i\left(|+1\rangle\langle 0| - |0\rangle\langle +1|\right)$$

$$\hat{\lambda}_7 = -i|0\rangle\langle -1| + h.c.$$

where $\hat{\lambda}_2, \hat{\lambda}_7$ – operators described by Gell-Mann matrices, $\hat{E}$ – identity operator. Here transverse hyperfine interaction is neglected as $A_\perp \ll D$.

To simulate the experiment, described by the sequence in Figure 1a, let us start from the mixed state, described by the density matrix $\rho(0)$:

$$\rho(0) = |0\rangle\langle 0| \otimes \frac{1}{3}\left(|+1\rangle\langle +1| + |0\rangle\langle 0| + |-1\rangle\langle -1|\right) , \quad (6)$$

where it is supposed that the laser pulse before MW pulses fully polarized the electronic spin into the $m_S = 0$ substate and the nuclear spin has a probability distribution near its thermal equilibrium. The density matrix of the system state after MW pulses $\rho(t_{MW})$ is obtained by the numerical solution of the Lindblad equation:

$$\dot{\rho}(t) = -\frac{i}{\hbar}\left[H'_{MW}(t) + H'_0, \rho(t)\right] \quad (7)$$

After MW pulse application, RF pulses should be applied to the flip nuclear spin on $m_S = \pm 1$ manifolds. The resulting flip is simulated by applying the evolution operator described as:

$$\hat{U}_{RF} = |0\rangle\langle 0| \otimes E + |+1\rangle\langle +1| \otimes \left(i\big[|+1\rangle\langle 0| + |0\rangle\langle +1|\big] + |-1\rangle\langle -1|\right) +$$
$$+ |-1\rangle\langle -1| \otimes \left(i\big[|-1\rangle\langle 0| + |0\rangle\langle -1|\big] + |+1\rangle\langle +1|\right) \quad ,\quad (8)$$

resulting in $\rho_{final} = \hat{U}_{RF}\rho(t_{MW})\hat{U}^+_{RF}$. In the sequence under consideration the RF pulse is followed by a light pulse that repopulates the $m_S = 0$ substate. It is acknowledged that this process does not materially alter the nuclear spin as the nuclear state can withstand numerous light pulses. Here such influence is neglected. The resulting population of $m_I = 0$ is estimated to be $P_{m_I=0} = \langle 0|Tr_e \rho_{final}|0\rangle$.

To simulate inhomogeneous broadening due to the interaction of an ensemble of the $NV^-$ center with surrounding impurities, it is assumed that the magnetic field $B_z$ has the Cauchy distribution with the maximum of probability density at the external field $B_0$ and standard deviation, corresponding to the width of the experimentally observed ODMR line $\gamma_B = \Delta_{FWHM}/\gamma_e$. The authors formed 201 equally spaced ensembles with magnetic fields $\{B_z^i\}_{i \in [0,200]}$ within the $[B_0 - 6\Delta_{FWHM}/\gamma_e; 6\Delta_{FWHM}/\gamma_e]$ range and solved the aforementioned Lindblad equation for each ensemble, leading to a set of solutions for the $m_I = 0$ substate population: $\{P_{m_I=0}^i\}_i$. The ensemble-averaged $m_I = 0$ state population is obtained by convolution with the Cauchy distribution:

$$P_{m_I=0}^{avg} = \frac{\sum_i P_{m_I=0}^i w_i}{\sum_i w_i}, \qquad (9)$$

where $w_i = 1/\left((B_z^i - B_0)^2 + \gamma_B^2/4\right)$ is the ensemble weight in the distribution, $P_{m_I=0}^{avg}$ is target for maximization when the optimal pulse envelope and detuning are searched for.

In the current work, 3 types of pulse shapes are investigated: square, Gaussian and band-selective Shinnar-Le Roux [31] (see Figure 1b).

The square pulse is the simplest form of the excitation waveform. It can be easily implemented experimentally by using on-off modulation with an MW switch. However, the excitation profile (Figure 2a) of the square pulse has a series of side lobes, which will lead to the excitation of magnetic transitions conditioned by undesired nuclear spin projection values. It is possible to tune Rabi frequency and pulse duration in a way that neighbor resonance will coincide with zero excitation. If detuning to neighbor resonance is equal to $|A_\parallel|$, as it is for the $NV^-$ center, then the $\pi$-pulse with $\Omega_{Rabi} = |A_\parallel|/\sqrt{3}$ will meet this condition (see Appendix). It was discovered that the found Rabi frequency for the optimal square pulse was close to the aforementioned value. To run the optimization algorithm, the authors parametrized Rabi frequency, durations, and detunings for both MW pulses and the maximized $P_{m_I=0}^{avg}$ value.

The Gaussian pulse does not have side lobes (see Figure 2b), but requires an arbitrary waveform generator to be included in the setup. For maximization, the authors parametrized the detuning and amplitude of the gauss impulse, assuming it is $\pi$ pulse for zero detuning.

For the Shinnar-Le Roux band-selective pulse (see Figure 2c), the authors used the procedure described in the work [31] with a pulse length of 4 microseconds and a pulse bandwidth of 4 MHz. Only the detuning of the algorithm was parametrized.

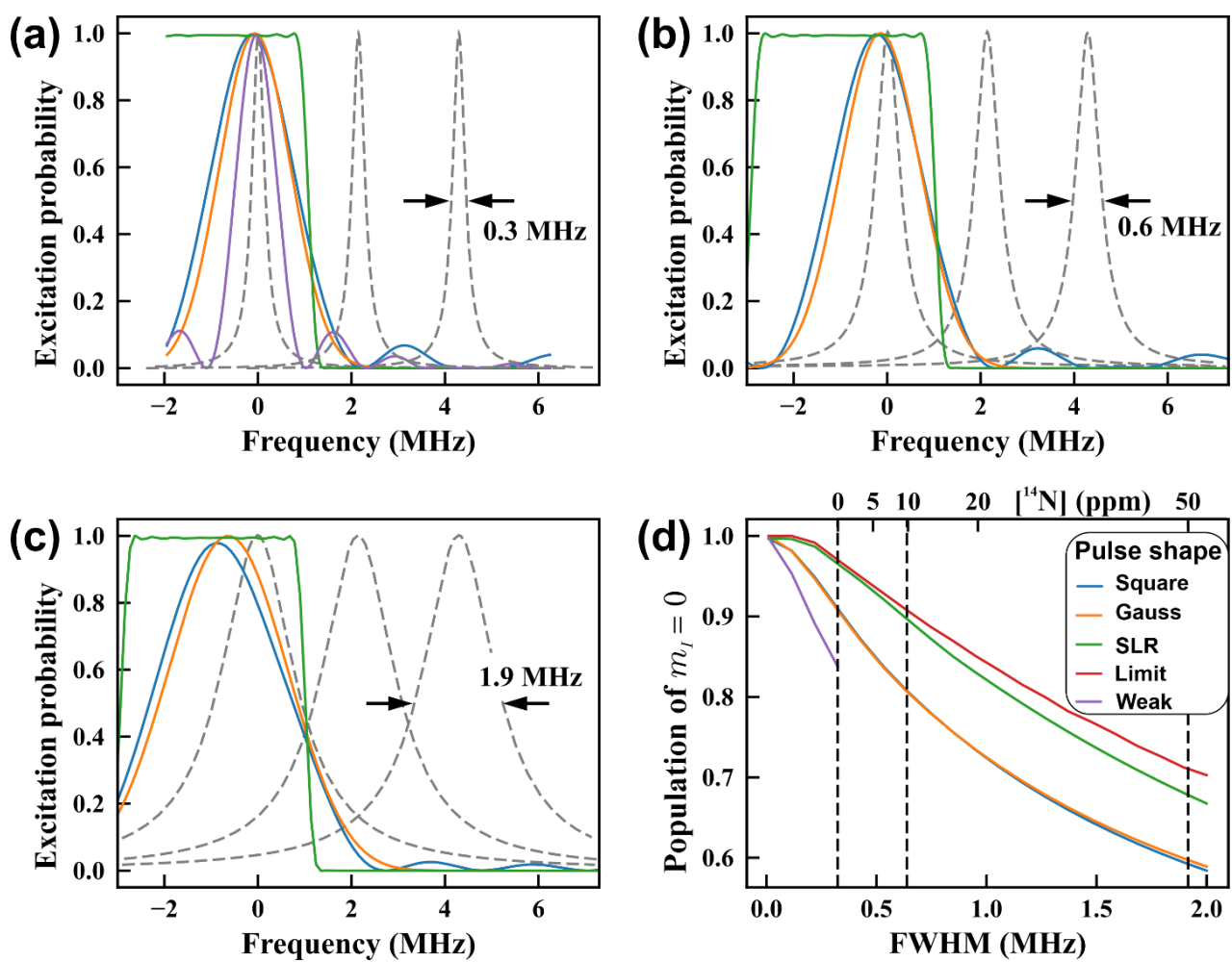

Figure 2 a, b, c) – solid lines correspond to excitation profiles, dashed lines show resonance contours of the $NV^-$ center. Pulses and profiles are shown for the first pulse in the DNP sequence. d) The quality of the nuclear spin state preparation using the DND sequence for several pulse shapes. The green curve stands for theoretical limit, orange – for optimized squire shape pulses, blue – for optimized frequency-selective pulses.

To compare the performance of all pulse shapes, the authors simulated the boundary of possible population of $m_I = 0$ state for specific transition linewidth. For the pulse acting of $|m_s = 0; m_I = -1\rangle \leftrightarrow |m_s = -1; m_I = -1\rangle$ transition, if transition frequency is equal to lower than $f_{-1} + |A_\parallel|/2$, the population is fully inversed. If transition frequency is higher, no inversion occurs at all.

Optimal values for pulses can be found in Table 1. The resulting performance of the pulse application is depicted in Figure 2d, showing that the Shinnar-Le Roux pulse performs almost at the ultimate limit. At the same time, the Gaussian pulse performs equally well as a simpler square pulse. The increase in the polarization efficiency provided by the Shinnar-Le Roux pulse nevertheless strongly depends on the width of the magnetic transition, which, according to [24], is closely related to the nitrogen concentration in the sample. Indeed, if the full width half maximum (FWHM) of the magnetic dipole transition is smaller than ≈350 kHz, typical for low nitrogen content samples [24], the increase in the population of the target ($m_I = 0$) state from the use of the optimized pulse does not exceed 6% in state purity. At the same time, if samples with high content of nitrogen are utilized [14], the increase in the population of the target state from the utilization of the optimized pulse can be as large as 15% to FWHM of 1.5 MHz. While the benefits from the usage of the optimized pulse continue to increase at higher widths, the degree of nuclear spin polarization quite expectedly goes down. Thus, even an optimized pulse cannot prevent degradation of the nuclear polarization in the event of a strong overlap of the broadened magnetic dipole transitions, but it can provide a significant improvement of the polarization process halfway toward this overlap. It is important to note that a high nitrogen content sample may show better performance for magnetometry application due to higher nitrogen to $NV^-$ conversion and smaller parasitic absorption of the laser excitation by impurities other than $NV^-$ [32]. Thus, the use of these kinds of samples may be quite interesting for sensors using a nuclear spin if the optimized pulse sequence described above is utilized.

Table 1 Optimal pulse parameters.

|  | | Linewidth, MHz | | | | | | | | | |
|---|---|---|---|---|---|---|---|---|---|---|---|
|  | | 0.01 | 0.15 | 0.32 | 0.43 | 0.64 | 0.95 | 1.27 | 1.48 | 1.79 | 2.00 |
|  | $F_{Rabi}^{(-1)}$, MHz | 1.13 | 1.14 | 1.16 | 1.18 | 1.20 | 1.25 | 1.32 | 1.37 | 1.48 | 1.61 |

|  |  |  |  |  |  |  |  |  |  |  |
|---|---|---|---|---|---|---|---|---|---|---|
| Square pulse | $F_{Rabi}^{(+1)}$, MHz | 1.24 | 1.27 | 1.34 | 1.38 | 1.44 | 1.57 | 1.70 | 1.80 | 1.96 | 2.07 |
| | $\delta_{-1}$, MHz | 0.03 | 0.01 | −0.04 | −0.06 | -0.10 | -0.18 | -0.27 | -0.34 | -0.46 | -0.57 |
| | $\delta_{+1}$, MHz | 0.00 | -0.03 | -0.09 | -0.13 | -0.19 | -0.30 | -0.44 | -0.53 | -0.69 | -0.80 |
| | $\Delta T_{MW}^{(-1)}$, % | 1.1 | 1.4 | 1.9 | 2.2 | 2.5 | 3.0 | 3.5 | 3.8 | 4.3 | 4.6 |
| | $\Delta T_{MW}^{(+1)}$, % | 0.0 | 0.0 | 0.1 | 0.1 | 0.1 | 0.0 | 0.0 | -0.1 | -0.1 | -0.2 |
| Gauss pulse | $F_{Rabi}^{(-1)}, F_{Rabi}^{(+1)}$, MHz | 1.00 | 1.27 | 1.42 | 1.48 | 1.58 | 1.74 | 1.93 | 2.06 | 2.28 | 2.45 |
| | $\delta_{+1}, \delta_{-1}$, MHz | 0.00 | -0.02 | -0.06 | -0.09 | -0.14 | -0.24 | -0.36 | -0.44 | -0.59 | -0.69 |
| Shinnar-Le Roux | $\delta_{+1}, \delta_{-1}$, MHz | -0.84 | -0.87 | -0.89 | -0.89 | -0.95 | -0.94 | -0.94 | -0.95 | -0.96 | -0.96 |
| Polarization | Square pulse | 0.997 | 0.97 | 0.91 | 0.87 | 0.81 | 0.73 | 0.68 | 0.64 | 0.61 | 0.58 |
| | Gaussian pulse | 1.00 | 0.97 | 0.91 | 0.87 | 0.81 | 0.73 | 0.68 | 0.65 | 0.61 | 0.59 |
| | Shinnar-Le Roux | 1.00 | 1.00 | 0.97 | 0.94 | 0.90 | 0.83 | 0.77 | 0.74 | 0.69 | 0.67 |
| | Limit | 1 | 1 | 0.97 | 0.95 | 0.91 | 0.85 | 0.8 | 0.77 | 0.73 | 0.7 |

## IV. CONCLUSION

The Shinnar-Le Roux algorithm was used to generate an optimized shape of the pulse in the dynamic nuclear polarization sequence for the ground state of the $NV^-$ center in diamond. The sequence was composed of two consequent $\pi$ pulses, the shape of which was optimized, accompanied by light and radiofrequency pulses. In the case of dense ensembles of $NV^-$, it was found that the Shinnar-Le Roux pulse shape with optimized parameters gave a significant increase in performance compared to the simple square pulse or more advanced Gaussian shape pulse. More specifically, it was found that for 0.64 MHz FWHM of magnetic dipole transition, corresponding to a 10 ppm nitrogen concentration in diamond, the increase in the population of the target state could reach up to a factor of 1.11, while in the case of 1.9 MHz FWHM corresponding to about 50 ppm nitrogen concentration, the increase could exceed a factor of 1.15. This work thus paves the way to the efficient use of high nitrogen density diamond samples, already acknowledged in magnetometry, for the nuclear spin utilizing sensor based on $NV^-$ in diamond.

## V. ACKNOWLEDGMENTS

This study was supported by Russian Science Foundation grant No. 21-42-04407. The authors acknowledge support from Rosatom.

## APPENDIX

To avoid cross-talk between two close frequency two-level systems, it is possible to select Rabi frequency so that it makes $\pi$ rotation for one system and $2\pi$ for another, thus the first one is inverted, and the second one is untouched [33]. If one of the systems is in resonance and its Rabi frequency is $\Omega_R$ and the other one responds to the electromagnetic wave in the same way but is detuned on $\delta\omega$, then for the second one, the Rabi frequency would be $\Omega = \sqrt{\Omega_R^2 + \delta\omega^2}$. Assuming square shape pulses, if one wants to have $\pi$ rotation for one system and $2\pi$ for another, then the pulse of the length $\tau$ must satisfy to:

$$\begin{aligned}\Omega_R \tau &= \pi \\ \Omega \tau &= 2\pi\end{aligned} \quad (10)$$

This leads to:

$$2\Omega_R = \Omega = \sqrt{\Omega_R^2 + \delta\omega^2}, \quad (11)$$

and, finally:

$$\Omega_R = \delta\omega / \sqrt{3}. \quad (12)$$